\documentclass{PoS}

\title{General Theoretical Introduction to Hadronic $B$ Decays}

\ShortTitle{General Theoretical Introduction}

\author{\speaker{Hsiang-nan Li}
\\
        Institute of Physics, Academia Sinica, Nankang,
Taipei 115, Taiwan, Republic of China\\
        E-mail: \email{hnli@phys.sinica.edu.tw}}

\abstract{I briefly introduce the theoretical frameworks for studies
of two-body hadronic $B$ meson decays, which include the
factorization assumption, the QCD-improved factorization, the
perturbative QCD, the soft-collinear effective theory, the
light-cone QCD sum rules, and the quark-diagram parametrization.}

\FullConference{Flavor Physics and CP Violation 2009\\
         May 27 - June 1, 2009\\
         Lake Placid, NY, USA}

\begin{document}

\section{Introduction}

Hadronic $B$ meson decays are difficult to analyze because of
complicated QCD dynamics and multiple characteristic scales they
involve: the $W$ boson mass $m_W$, the $b$ quark mass $m_b$, and the
QCD scale $\Lambda_{\rm QCD}$. The standard procedure is first to
integrate out the scale $m_W$, such that QCD dynamics is organized
into an effective weak Hamiltonian \cite{REVIEW}. For the $B\to
D\pi$ decays, the effective Hamiltonian is written as
\begin{eqnarray}
{\cal H}_{\rm eff} = {G_F\over\sqrt{2}}\, V_{cb}V_{ud}^*
\Big[C_1(\mu)O_1(\mu)+C_2(\mu)O_2(\mu)\Big]\;,
\end{eqnarray}
where $G_F$ is the Fermi coupling constant, $V_{cb}V_{ud}^*$ is the
product of the Cabibbo-Kobayashi-Maskawa matrix elements, $\mu$ is
the renormalization scale, $C_{1,2}$ are the Wilson coefficients,
and the four-fermion operators are defined by
\begin{eqnarray}
O_1 = (\bar db)_{V-A}(\bar cu)_{V-A}\;,\qquad\qquad O_2= (\bar
cb)_{V-A}(\bar du)_{V-A}.
\end{eqnarray}

To derive $B\to D\pi$ decay amplitudes, one evaluates the hadronic
matrix elements $\langle D\pi|O_i(\mu)|B\rangle$. Different
theoretical approaches have been developed for this evaluation,
which include the factorization assumption, the QCD-improved
factorization, the perturbative QCD, the soft-collinear effective
theory, the light-cone QCD sum rules, and the quark-diagram
parametrization. In this talk I briefly introduce the basic ideas of
these approaches \cite{L03}.

\section{Factorization Assumption}

Intuitively, decay products from a heavy $b$ quark move fast without
further interaction between them. This naive picture is supported by
the color-transparency argument \cite{transparency}: the Lorentz
contraction renders energetic final states emitted from the weak
vertex have small longitudinal color dipoles, which can not be
resolved by soft gluons. Therefore, the hadronic matrix element
$\langle O(\mu)\rangle$ is factorized into a product of two matrix
elements of single currents, governed by decay constants and form
factors, without soft gluon exchanges between them. This
factorization assumption (FA) \cite{BSW} was first proved in the
framework of large energy effective theory \cite{leet}, and
justified in the large $N_c$ limit \cite{largeN}, $N_c$ being the
number of colors. For the $B\to D\pi$ decays, the color-allowed
(color-suppressed) amplitude, involving the $B\to D$ ($B\to\pi$)
transition form factor, is proportional to the Wilson coefficient
$a_1=C_2+C_1/N_c$ ($a_2=C_1+C_2/N_c$).

In spite of its simplicity, the FA encounters three principal
difficulties. First, a hadronic matrix element under the FA is
independent of the renormalization scale $\mu$, as the vector or
axial-vector current is partially conserved. Consequently, the
amplitude $C(\mu)\langle O\rangle_{\rm fact}$ is not truly physical
as the scale dependence of the Wilson coefficient does not get
compensation from the matrix element. This problem may not be
serious for color-allowed modes, because the parameter $a_1$ is
roughly independent of $\mu$. It is then not a surprise that the
simple FA gives predictions in relatively good agreement with data
of these modes. However, the parameter $a_2$ depends strongly on the
renormalization scale and on the renormalization scheme, because of
the similar magnitude and different sign of the $C_1(\mu)$ and
$C_2(\mu)/N_c$ terms (calculated in the NDR scheme and for
$\Lambda_{\overline{MS}}^{(5)} = 225$ GeV, the Wilson coefficients
have the values $C_1(m_B) = -0.185$ and $C_2(m_B) = 1.082$
\cite{REVIEW}, $m_B$ being the $B$ meson mass). This may be the
reason the FA fails to accommodate data of color-suppressed modes.
It also means that $a_2$ is more sensitive to subleading
contributions.

The second difficulty is related to the first one: nonfactorizable
effects have been neglected in the FA. This neglect may be justified
for color-allowed modes due to the large and roughly
$\mu$-independent value of $a_1$, but not for color-suppressed
modes, such as $B\to J/\psi K^{(*)}$. The $J/\psi$ meson emitted
from the weak vertex is not energetic, and the color-transparency
argument does not apply. To circumvent this difficulty,
nonfactorizable contributions were parameterized into the parameters
$\chi_i$ \cite{Cheng94,Soares},
\begin{eqnarray}
a_1^{\rm eff}& =& C_2(\mu) + C_1(\mu) \left[{1\over N_c}
+\chi_1(\mu)\right]\,,
\nonumber\\
a_2^{\rm eff}& =& C_1(\mu) + C_2(\mu)\left[{1\over N_c} +
\chi_2(\mu)\right]\,.
\end{eqnarray}
The $\mu$ dependence of the Wilson coefficients is assumed to be
exactly compensated by that of $\chi_i(\mu)$ \cite{NRSX}. It is
obvious that the introduction of $\chi_i$ does not really resolve
the scale problem in the FA.

Third, strong phases are essential for predicting CP asymmetries in
exclusive $B$ meson decays. These phases, arising from the
Bander-Silverman-Soni (BSS) mechanism \cite{BSS}, are ambiguous in
the FA: the $c$ quark loop contributes an imaginary piece
proportional to
\begin{equation}
\int du u(1-u)\theta(q^2 u(1-u)-m_c^2)\;, \label{stp}
\end{equation}
where $q^2$ is the invariant mass of the gluon emitted from the
penguin. Since $q^2$ is not precisely defined in the FA, one can not
obtain definite information of strong phases from Eq.~(\ref{stp}).
Moreover, it is legitimate to question whether the BSS mechanism is
an important source of strong phases in $B$ meson decays. Viewing
the above difficulties, the FA is not a complete model, and it is
necessary to go beyond the FA by developing reliable and systematic
theoretical approaches.

\section{QCD-improved Factorization}

\begin{figure}
\begin{center}
\includegraphics[width=.6\textwidth]{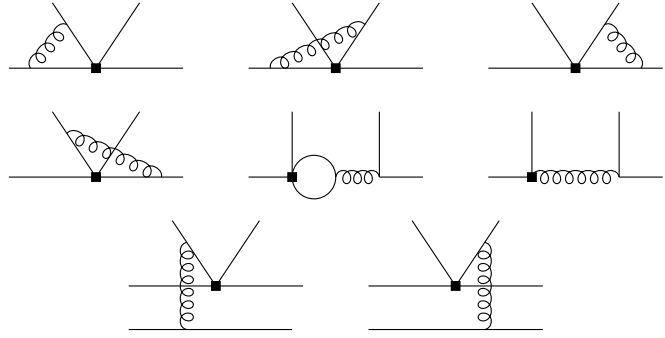}
\end{center}
\caption{$O(\alpha_s)$ corrections to the FA in the QCDF approach.}
\label{fig1}
\end{figure}

The color-transparency argument allows the addition of hard gluons
between the energetic mesons emitted from the weak vertex and the
$B$ meson transition form factors. These hard gluon exchanges lead
to higher-order corrections in the coupling constant $\alpha_s$ to
the FA. By means of Feynman diagrams, they appear as the vertex
corrections in the first two rows of Fig.~\ref{fig1} \cite{BBNS}. It
has been shown that soft divergences cancel among them, when
computed in the collinear factorization theorem. These $O(\alpha_s)$
corrections weaken the $\mu$ dependence in Wilson coefficients, and
generate strong phases. Besides, hard gluons can also be added to
form the spectator diagrams in the last row of Fig.~\ref{fig1}.
Feynman rules of these two diagrams differ by a minus sign in the
soft region resulting from the involved quark and anti-quark
propagators. Including the above nonfactorizable corrections to the
FA leads to the QCD-improved factorization (QCDF) approach
\cite{BBNS}. The gluon invariant mass $q^2$ in the BSS mechanism can
be unambiguously defined and related to parton momentum fractions in
QCDF. Hence, the theoretical difficulties in the FA are resolved.
This is an important step towards a rigorous framework for two-body
hadronic $B$ meson decays in the heavy quark limit.

\begin{figure}
\begin{center}
\includegraphics[width=.9\textwidth]{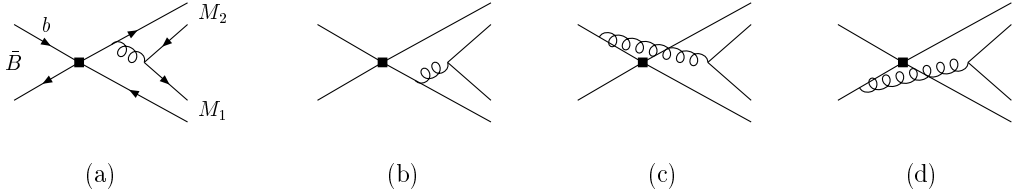}
\end{center}
\caption{Annihilation contributions.} \label{fig2}
\end{figure}

Corrections in higher powers of $1/m_b$ to the FA can also be
included into QCDF, such as those from the annihilation topology in
Fig.~\ref{fig2}, and from twist-3 contributions to the spectator
amplitudes. However, it has been found that endpoint singularities
exist in these high-power contributions, which arise from the
divergent integral $\int_0^1 dx/x$, $x$ being a momentum fraction.
Similar singularities are developed, when applying the collinear
factorization to $B$ meson transition form factors. Because of the
endpoint singularities, the annihilation and twist-3 spectator
contributions must be parameterized as \cite{BBNS}
\begin{eqnarray}
\ln\frac{m_B}{\Lambda_h}\left(1+\rho_Ae^{i\delta_A}\right)\;,\;\;\;\;
\ln\frac{m_B}{\Lambda_h}\left(1+\rho_He^{i\delta_H}\right)\;,
\label{rhoa}
\end{eqnarray}
respectively, with the hadronic scale $\Lambda_h$. A QCDF formula
then contains the arbitrary parameters $\rho_{A,H}$ and
$\delta_{A,H}$. Setting these parameters to zero, one obtains
predictions in the ''default" scenario, and the variation of the
arbitrary parameters gives theoretical uncertainties. If tuning
these parameters to fit data, one obtains results in the scenarios
''S1", ''S2",...\cite{BN03}.

\section{Perturbative QCD}

\begin{figure}
\begin{center}
\includegraphics[width=.6\textwidth]{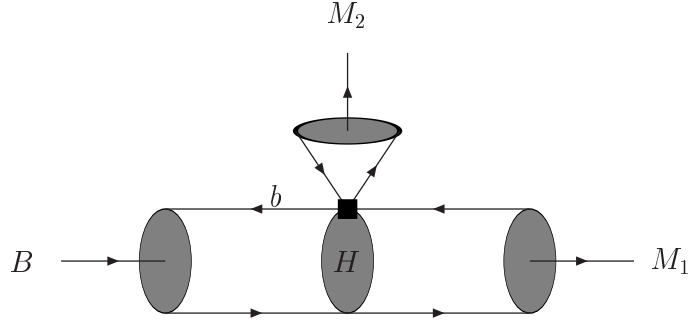}
\end{center}
\caption{Perturbative QCD factorization.} \label{fig3}
\end{figure}

The endpoint singularities signal the breakdown of the collinear
factorization for two-body hadronic $B$ meson decays. Motivated by
removing these singularities, the perturbative QCD (PQCD) approach
based on the $k_T$ factorization theorem was developed
\cite{LY1,CL,KLS,LUY}. A parton transverse momentum $k_T$ is
produced by gluon radiations, before hard scattering occurs. The
endpoint singularities from the small $x$ region indicate that the
parton transverse momentum $k_T$ is not negligible. Taking into
account $k_T$, a particle propagator does not diverge as $x\to 0$.
The $B$ meson transition form factors, and the spectator and
annihilation contributions are then all calculable in the framework
of the $k_T$ factorization. It has been shown that a $B\to M_1M_2$
decay amplitude is factorized into the convolution of the six-quark
hard kernel, the jet function and the Sudakov factor with the
bound-state wave functions as shown in Fig.~\ref{fig3},
\begin{eqnarray}
A(B\to M_1M_2)=\phi_B\otimes H\otimes J\otimes S
\otimes\phi_{M_1}\otimes \phi_{M_2}\;. \label{six}
\end{eqnarray}
The jet function $J$ comes from the threshold resummation, which
exhibits suppression in the small $x$ region \cite{THRE}. The
Sudakov factor $S$ comes from the $k_T$ resummation, which exhibits
suppression in the small $k_T$ region \cite{Botts:1989kf,Li:1992nu}.
Therefore, these resummation effects guarantee the removal of the
endpoint singularities. $J$ ($S$), organizing double logarithms in
the hard kernel (meson wave functions), is hidden in $H$ (the three
meson states) in Fig.~\ref{fig3}. The arbitrary parameters
introduced in QCDF \cite{BBNS} are not necessary, and PQCD involves
only universal and controllable inputs.

The theoretical difficulties in the FA are also resolved in PQCD but
in a different way. The FA limit of the PQCD approach at large
$m_b$, which is not as obvious as in QCDF, has been examined
\cite{THRE}. It was found that the factorizable emission amplitude
decreases like $m_b^{-3/2}$, if the $B$ meson decay constant $f_B$
scales like $f_B\propto m_b^{-1/2}$. This power-law behavior is
consistent with that obtained in \cite{BBNS,Chernyak:1990ag}. The
higher-order corrections to the FA have been included in PQCD, which
moderate the dependence on the renormalization scale $\mu$. The
ratio of the spectator contribution over the factorizable emission
contribution decreases with $m_b$ in PQCD, showing a behavior close
to that in QCDF. The gluon invariant mass $q^2$ in the BSS mechanism
is clearly defined and related to parton momentum fractions. The
penguin annihilation amplitude is almost imaginary in PQCD
\cite{KLS}, whose mechanism is similar to the BSS one \cite{BSS}: in
the annihilation topology, the loop is formed by the internal
particles in the LO hard kernel and by infinitely many Sudakov
gluons exchanged between two partons in a light meson. A sizable
strong phase is generated, when the internal particles go on mass
shell. In terms of the principle-value prescription for the internal
particle propagator, the strong phase is given by \cite{KLS}
\begin{eqnarray}
\frac{1}{xm_B^2-k_T^2+i\epsilon}=\frac{P}{xm_B^2-k_T^2}
-i\pi\delta(xm_B^2-k_T^2).
\end{eqnarray}

\section{Soft-Collinear Effective Theory}

\begin{figure}
\begin{center}
\includegraphics[width=.3\textwidth]{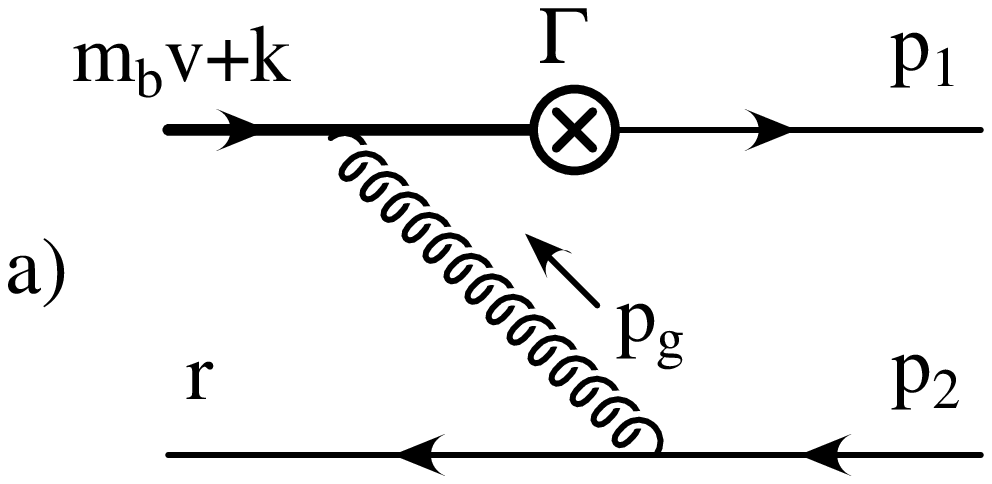}\hspace{1.0cm}
\includegraphics[width=.3\textwidth]{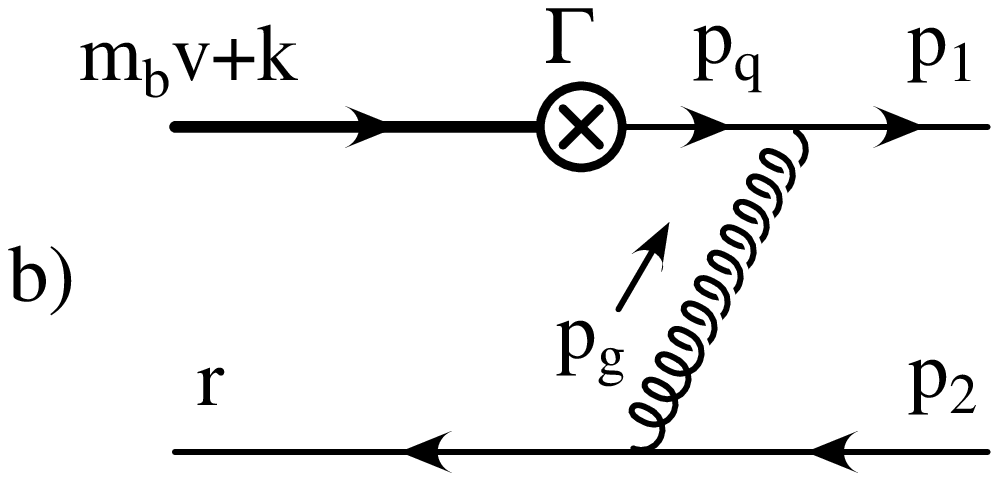}
\end{center}
\caption{Diagrams for the $B\to\pi$ form factor in QCD.}
\label{fig4}
\end{figure}

The soft-collinear effective theory (SCET) based on the collinear
factorization is formulated in the framework of operator product
expansion (OPE) \cite{bfl,bfps,cbis,bpssoft}. The matching at
different scales involved in $B$ meson decays has been carefully
handled in SCET. Take the simple $B\to\pi$ transition form factor in
Fig.~\ref{fig4} as an example. The soft spectator in the $B$ meson
carries the momentum $r\sim O(\Lambda_{\rm QCD})$, because it is
dominated by soft dynamics. If the spectator in the energetic pion
carries the momentum $p_2 \sim O(m_b)$, the virtual gluon in
Fig.~\ref{fig4} is off-shell by $p_g^2=(p_2-r)^2=-2p_2\cdot r \sim
O(m_b\Lambda_{\rm QCD})$. Then the virtual quark in
Figs.~\ref{fig4}(a) is off-shell by $(m_bv+k+p_g)^2-m_b^2\sim
O(m_b^2)$, where $v$ is the $b$ quark velocity and $k\sim
O(\Lambda_{\rm QCD})$ denotes the Fermi motion of the $b$ quark.
Hence, $B$ meson decays contain three scales below $m_W$: $m_b$,
$\sqrt{m_b\Lambda_{\rm QCD}}$, and $\Lambda_{\rm QCD}$.

\begin{figure}
\begin{center}
\includegraphics[width=.3\textwidth]{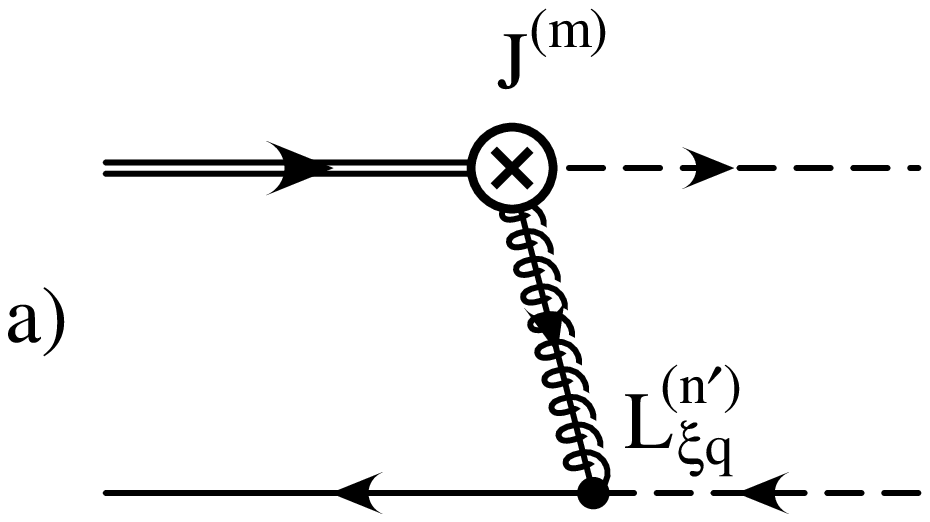}\hspace{1.0cm}
\includegraphics[width=.3\textwidth]{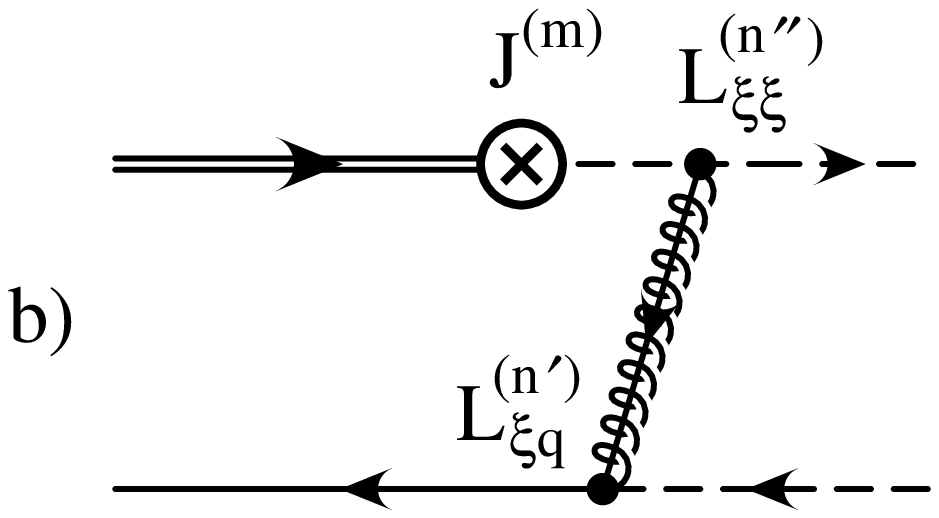}
\end{center}
\caption{Diagrams for the $B\to\pi$ form factor in SCET$_{\rm I}$.}
\label{fig5}
\end{figure}

\begin{figure}
\begin{center}
\includegraphics[width=.35\textwidth]{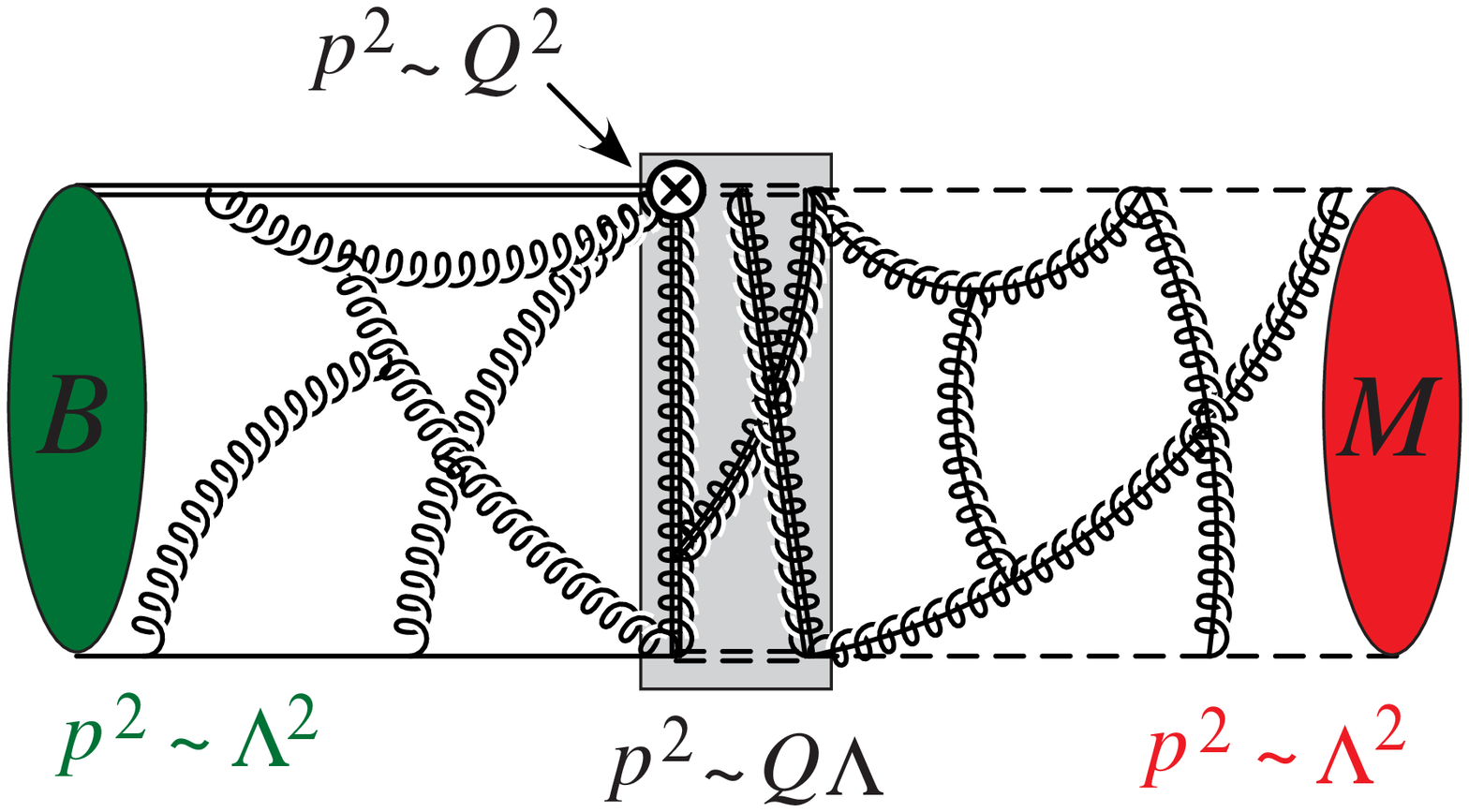}\hspace{1.0cm}
\includegraphics[width=.35\textwidth]{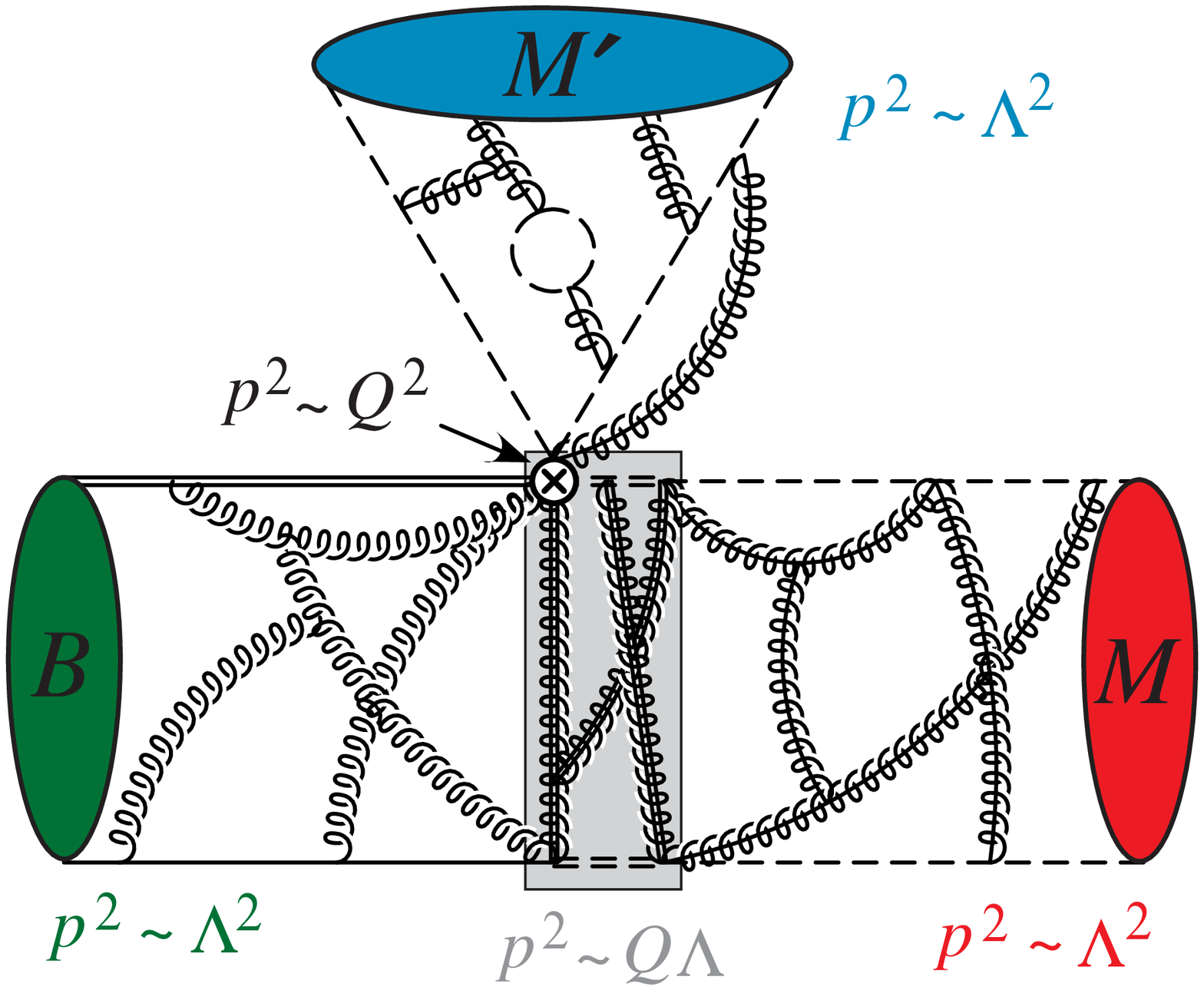}
\end{center}
\caption{Factorization of the $B\to\pi$ form factor and of the $B\to
M_1M_2$ decay in SCET.} \label{fig6}
\end{figure}

The separate matching at the two scales $m_b$ and
$\sqrt{m_b\Lambda_{\rm QCD}}$ is briefly explained below
\cite{BPS03}. The first step is to integrate out the lines off-shell
by $m_b^2$ in QCD, and the resultant effective theory is called
SCET$_{\rm I}$. One then derives the zeroth-order effective current
$J^{(0)}$ from the $b\to u$ weak vertex, and the first-order
effective current $J^{(1)}$ by shrinking the virtual $b$ quark line
in Fig.~\ref{fig4}(a). The next step is to integrate out the lines
off-shell by $m_b\Lambda_{\rm QCD}$ in SCET$_{\rm I}$, arriving at
SCET$_{\rm II}$. The relevant diagrams to start with are displayed
in Fig.~\ref{fig5}. Shrinking all the lines off-shell by
$m_b\Lambda_{\rm QCD}$, one derives the corresponding Wilson
coefficients, i.e., the jet functions, and the effective
four-fermion operators. Sandwiching these four-fermion operators by
the initial $B$ meson state and the final pion state leads to the
$B$ meson and pion distribution amplitudes. The $B\to\pi$ transition
form factor is then factorized as depicted in Fig.~\ref{fig6}. The
factorization of two-body hadronic $B$ meson decays is constructed
in a similar way, and the result is also shown in Fig.~\ref{fig6}.

At leading power in $1/m_b$, there is no large source of strong
phases in SCET (the annihilation contribution is parametrically
power-suppressed). To acquire strong phases, it has been argued that
$c\bar c$ (charming) penguins could give long-distance effects at
leading power \cite{BPRS04}. This contribution is nonperturbative,
so it must be parameterized as an arbitrary amplitude $A^{c\bar c}$.
Including the charming penguin, SCET has been applied as an
QCD-improved parametrization, and $A^{c\bar c}$ is determined
together with other hadronic inputs from data. It should be
mentioned that the long-distance charming-penguin contribution is
power-suppressed according to QCDF, PQCD and light-cone sum rules
\cite{KMM03}.

\section{Light-Cone Sum Rules}

QCD sum rules \cite{SVZ,ISNR} are based on the quark-hadron duality,
which is very different from a factorization theorem. A simple
argument of the quark-hadron duality has been presented in
\cite{CRS93,KR}: consider evaluation of a correlation function by
means of the dispersion relation. One can choose either a contour
along the real axis, which may be close to a physical pole, or a
contour far away from the physical pole. When moving along the
former contour, one picks up nonperturbative contributions to the
correlation function from the pole. When moving along the latter,
the correlation function can be evaluated in the framework of OPE.
If there is no other pole inside the combined contour of the former
and the latter, the above two choices should give the same result.
This explains the idea of the quark-hadron duality.

QCD sum rules have been applied to various problems in heavy flavor
physics. Take the $B$ meson decay constant $f_B$ as an example
\cite{6auth,RRY,fB}, which is defined via the matrix element
$\langle 0| m_B\bar{q}i\gamma_5 b| \bar B\rangle=f_Bm_B^2$, $q=u,d$.
Start with the correlation function of two heavy-light currents,
\begin{eqnarray}
\Pi(q^2)=i \int d^4ye^{iq\cdot y}\langle 0| T[m_B\bar{q}i\gamma_5
b(y), m_B\bar{b}i\gamma_5 q(0)]| 0\rangle\;, \label{fBcorr}
\end{eqnarray}
which can be treated by OPE at the quark level, if $q^2$ is far
below $m_b^2$, or parameterized as a sum over hadronic states
including the ground-state $B$ meson for $q^2 \geq m_B^2$. The
quark-hadron duality relates the expressions in these two regions.
Therefore, on the left-hand (hadron) side of the sum rule, one has
\begin{eqnarray}
\Pi(q^2)= \frac{f_B^2m_B^4}{m_B^2-q^2}+\cdots\;,
\label{dispfB}
\end{eqnarray}
where the contribution of the ground-state $B$ meson has been
singled out, and $\cdots$ represents those from the excited
resonances and from the continuum of hadronic states with the $B$
meson quantum numbers. On the right-hand (quark) side of the sum
rule, one has the expansion including the perturbative series in
$\alpha_s$ and the quark, gluon and quark-gluon condensates.
Evaluating the right-hand side of the sum rule, one is able to
estimate $f_B$.

Light-cone sum rules (LCSR) \cite{lcsr} is a simplified version of
QCD sum rules. Consider the $B\to \pi $ transition form factors
\cite{KR,PB3,BKR}, for which the correlation function is chosen as
\begin{eqnarray}
i \int \!d^4ye^{iq\cdot y}\langle \pi^+(P_2)|
T[\bar{u}\gamma_\lambda b(y), m_b\bar{b}i\gamma_5 d(0)]| 0\rangle,
\label{lcsrcorr}
\end{eqnarray}
with the current $\bar{u}\gamma_\lambda b(y)$ representing the weak
vertex. Compared to Eq.~(\ref{fBcorr}), the final state has been
specified as a pion, and the twist expansion has been applied to the
local current associated with the pion. LCSR has been extended to
the analysis of two-body hadronic $B$ meson decays in \cite{K00},
and interesting observations were obtained.

\section{Quark-diagram Parametrization}

\begin{figure}
\begin{center}
\includegraphics[width=.6\textwidth]{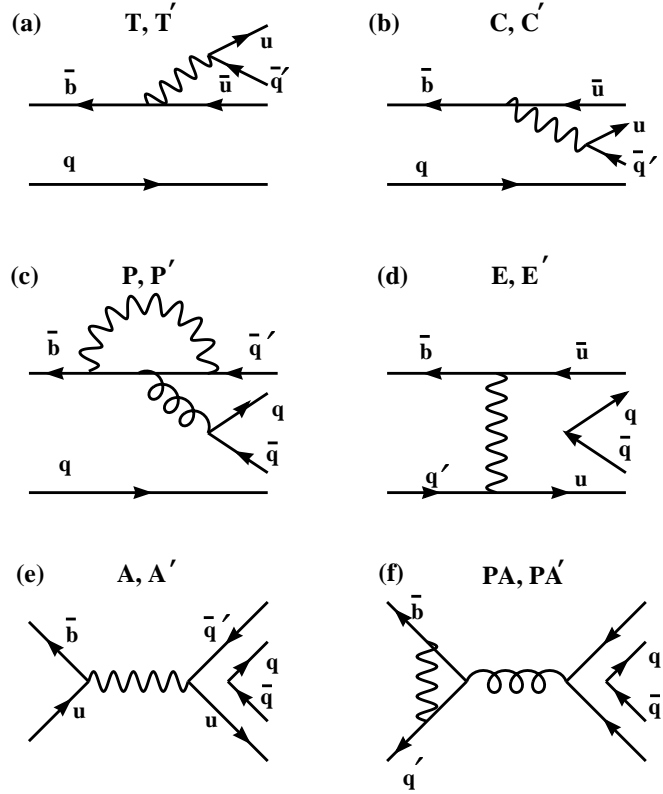}
\end{center}
\caption{Quark diagrams for the $B\to M_1M_2$ decays.} \label{fig7}
\end{figure}

The quark-diagram parametrization is a widely adopted approach to
two-body hadronic $B$ meson decays \cite{CC}. Various quark diagrams
are defined in Fig.~\ref{fig7}, with the color-allowed tree
amplitude $T$, the color-suppressed tree amplitude $C$, the penguin
amplitude $P$, the $W$-exchange amplitude $E$, the annihilation
amplitude $A$, and the penguin annihilation amplitude $PA$. One also
defines the electroweak penguin amplitude $P_{ew}$ and the
color-suppressed electroweak penguin amplitude $P_{ew}^c$. According
to the above definitions, the quark-diagram parametrization for the
$B\to\pi\pi$ decays is given by
\begin{eqnarray}
\sqrt{2}A(B^+\to \pi^+\pi^0)&=&-T\left[1+\frac{C}{T}
+\frac{P_{ew}}{T}e^{i\phi_2}\right]\;,
\nonumber\\
A(B_d^0\to \pi^+\pi^-)&=&-T\left(1 +\frac{P}{T}e^{i\phi_2}\right)\;,
\nonumber\\
\sqrt{2}A(B_d^0\to \pi^0\pi^0)&=&T\left[\left(
\frac{P}{T}-\frac{P_{ew}}{T}\right)
e^{i\phi_2}-\frac{C}{T}\right]\;, \label{Mbpi1}
\end{eqnarray}
with the weak phase $\phi_2$. The parametrization for other decays
can be written down in a similar way.

Predictive power of this approach arises from flavor symmetry, such
as $SU(3)$ \cite{GRL} and $U$-spin \cite{FGR}, which relate the
amplitudes among relevant modes. For example, the former relates the
color-allowed tree amplitudes with the light quark $q=u,d,s$ and
with the light quark $q'=d,s$ in Fig.~\ref{fig7}(a). The latter
relates the $B_d^0\to K^+\pi^-$ and $B_s^0\to\pi^+ K^-$ decay
amplitudes, and the $B_s^0\to K^+K^-$ and $B_d^0\to\pi^+\pi^-$ decay
amplitudes. Hence, one can determine the quark amplitudes (together
with the weak phases sometimes) from data of some modes, and then
use them to make predictions for other modes. However, it is
difficult to estimate symmetry breaking effects in this approach
\cite{GHLR}.

\section{Summary}

We have been able to go beyond the factorization assumption by
including QCD corrections. Different approaches have been developed:
in QCDF the higher-order corrections are computed in the collinear
factorization, but the high-power corrections must be parameterized
due to the existence of the endpoint singularities. There are no
endpoint singularities in PQCD, which is based on the $k_T$
factorization, and in SCET, which employs the zero-bin subtraction
\cite{MS06}. Therefore, $B$ meson transition form factors and
$1/m_b$ power-suppressed contributions are calculable in both
approaches. The difference is that SCET involves more arbitrary
parameters, such as the charming penguins, which are the main source
of strong phases. The annihilation contribution is the main source
of strong phases in PQCD. Since external lines are off-shell in the
dispersion relation on the OPE side of QCD sum rules or LCSR, a soft
contribution has a definition different from those in factorization
theorems. Hence, the dominance of the soft contribution in the
former does not apply to the latter. At last, predictive power of
the factorization approaches comes from universality of
nonperturbative inputs, such as meson wave functions, but that of
the quark-diagram parametrization comes from flavor symmetry.

This work was supported by the National Center for Theoretical
Sciences and National Science Council of R.O.C. under Grant No.
NSC-95-2112-M-050-MY3.

\end{document}